\documentclass[a4paper,11pt]{article}

\usepackage{booktabs}

\usepackage{lineno}
\usepackage[colorlinks=false]{hyperref}
\usepackage{bm}
\usepackage{graphicx}
\usepackage{amssymb}
\usepackage{amsmath}
\usepackage{tabularx}
\usepackage{natbib}
\usepackage{floatrow}
\usepackage{caption}
\usepackage{multirow}
\usepackage{amsmath}
\usepackage{amsfonts}
\usepackage{amssymb}
\usepackage{xcolor}
\usepackage{subfig}
\usepackage{graphicx,amsmath,bm}
\usepackage[binary-units=true]{siunitx}
\usepackage{array}
\usepackage{authblk}
\usepackage{rotating}
\usepackage{enumerate}
\usepackage{ragged2e}

\usepackage[left=15mm,right=15mm,top=1.5cm,bottom=1.5cm,includeheadfoot]{geometry}
\setcitestyle{square,numbers}
\begin{document}
	
	\title{Field Reconstruction for High-Frequency Electromagnetic Exposure Assessment Based on Deep Learning}
	
	\author[1]{Miao~Cao}
	\author[1]{Zicheng~Liu}
	\author[2]{Bazargul~Matkerim}
	\author[1]{Tongning~Wu}		
	\author[1]{Changyou~Li}
	\author[1]{Yali~Zong}
	\author[1]{Bo~Qi}

	\affil[1]{\scriptsize Department of Electronic Engineering, School of Electronics and Information, Northwestern Polytechnical University, Xi'an 710029, China}
	\affil[2]{\scriptsize Department of Computer Science, Al-Farabi Kazakh National University, Almaty 050040, Kazakhstan}
	\maketitle
	
	\abstract{
		Fifth-generation (5G) communication systems, operating in higher frequency bands from 3 to 300 GHz, provide unprecedented bandwidth to enable ultra-high data rates and low-latency services. However, the use of millimeter-wave frequencies raises public health concerns regarding prolonged electromagnetic radiation (EMR) exposure. Above 6 GHz, the incident power density (IPD) is used instead of the specific absorption rate (SAR) for exposure assessment, owing to the shallow penetration depth of millimeter waves. This paper proposes a hybrid field reconstruction framework that integrates classical electromagnetic algorithms with deep learning to evaluate the IPD of wireless communication devices operating at 30 GHz, thereby determining compliance with established RF exposure limits. An initial estimate of the electric field on the evaluation plane is obtained using a classical reconstruction algorithm, followed by refinement through a neural network model that learns the mapping between the initial and accurate values. A multi-antenna dataset, generated via full-wave simulation, is used for training and testing. The impacts of training strategy, initial-value algorithm, reconstruction distance, and measurement sampling density on model performance are analyzed. Results show that the proposed method significantly improves reconstruction accuracy, achieving an average relative error of 4.57\% for electric field reconstruction and 2.97\% for IPD estimation on the test dataset. Additionally, the effects of practical uncertainty factors, including probe misalignment, inter-probe coupling, and measurement noise, are quantitatively assessed. }
	
	\section{Introduction}
	%
	%
	%
	%
	By leveraging both centimetric (3–30 GHz) and millimetric (30–300 GHz) wave bands \cite{5Gmm-WaveAntenna}, fifth-generation (5G) wireless communication offers substantially wider channel bandwidths compared to previous generations, enabling ultra-high data rates and low-latency communication. Nevertheless, the transition to higher frequency bands has intensified concerns about the potential health risks associated with prolonged electromagnetic radiation (EMR) exposure \cite{IEEEC95}. The widespread deployment of base stations, access points, and user equipment operating in the millimeter-wave range increases EMR exposure levels in close proximity to users, making accurate characterization of exposure essential.
	
	At frequencies above 6 GHz, the specific absorption rate (SAR) is no longer the preferred metric for exposure assessment because of the extremely shallow penetration depth of electromagnetic waves. Instead, the absorbed power density (APD) is defined as the basic restriction, representing the electromagnetic power absorbed per unit area within biological tissue. Since direct measurement of APD is challenging in practice, the incident power density (IPD), referring to the power density of incident field on a surface in free space, is adopted as the corresponding reference level, the limit of which is set as 10 W/m² for the general public and 50 W/m² for occupational environments \cite{IEEEC95,ICNIRP}.
	
	According to IEC/IEEE 63195 \cite{63195–1,63195–2}, compliance evaluation requires assessing the IPD in the ultra-near region (minimum 2 mm) of the device under test (DUT). However, current field probes cannot effectively measure the vector electric field in this region due to their physical size and coupling effects with the DUT. This limitation motivates the need to reconstruct the electromagnetic field distribution in the ultra-near region from measurements taken further away.
	
	Near–far field transformation techniques, widely used in antenna measurements \cite{near/far-field1,near/far-field2}, can calculate the far-field pattern from near-field measurements \cite{equivalentcurrent1, equivalentcurrent2, PWEM1, PWEM2}, and conversely, reconstruct near-field distributions from far-field data to evaluate IPD \cite{IPD-PWE1,IPD-PWE2,IPD-ISM1,IPD-ISM2}. The plane wave expansion method (PWEM) expresses the electric field as a superposition of plane waves in multiple propagation directions \cite{PWE1,LIU,liu2017electromagnetic}. In IPD evaluation, the measured plane field is transformed into its spatial spectrum via a fast Fourier transform (FFT), corrected for propagation attenuation \cite{PWE2,PWE3}, and then inverse transformed to yield the reconstructed field. For instance, Sasaki et al. \cite{IPD-PWE1} applied PWEM to estimate the peak spatially averaged power density 0.15$\lambda$ from an antenna with a reconstruction error below 0.35 dB (8.4\%). Nevertheless, the finite integration region in Fourier transform introduces edge discontinuities, causing ripples and errors in the reconstruction.
	
	The inverse source method (ISM) \cite{ISM1,ISM2,ISM3} provides another transformation approach, relating the radiated field to equivalent sources via dyadic Green’s functions. Omi et al. \cite{IPD-ISM2} applied ISM with sparse measurement data, achieving psIPD reconstruction errors below 0.4 dB (9.6\%). However, solving the inverse integral equation for displacement currents is computationally demanding, and both PWEM and ISM suffer from ill-conditioned problems, \emph{i.e.}, small measurement errors can cause large deviations in reconstructed fields.
	
	To address these limitations, this study proposes a hybrid field reconstruction framework that combines classical algorithms with deep learning to mitigate ill-conditioning and enhance reconstruction stability. Using full-wave simulated datasets covering multiple antenna types, we investigate the influence of training strategies, initialization algorithms, reconstruction distance, and sampling density on performance. Measurement uncertainty factors, including probe positioning errors, inter-probe coupling, and measurement noise, are also quantified to assess robustness. This integration of classical electromagnetic computation and data-driven learning aims to deliver accurate and stable millimeter-wave exposure assessment, supporting efficient and reliable high-frequency RF safety characterization.
	
	\section{Method}
	\subsection{Incident Power Density}
	\begin{figure}[!t]
		\centering
		\includegraphics[width=3.3in]{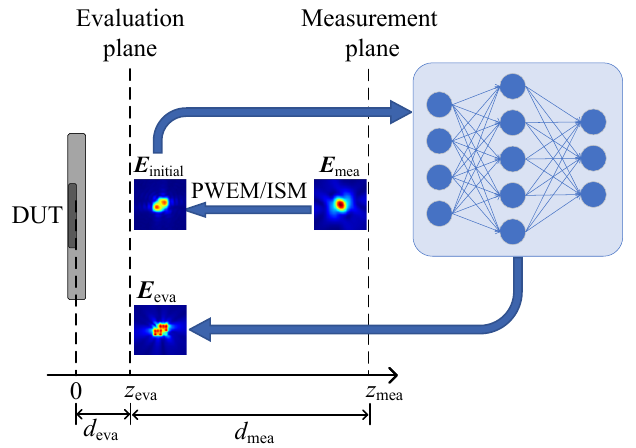}
		\caption{Schematic of the proposed field reconstruction method.}
		\label{principal}
	\end{figure}
	
	The proposed IPD evaluation framework is illustrated in Fig. 1. The evaluation plane is located at a distance $d_{\mathrm{eva}}$ from the device under test (DUT), and $d_{\mathrm{mea}}$ denotes the distance between the evaluation and measurement planes. Field reconstruction is performed in two stages.
	
	First, the electric field measured on the measurement plane is processed using the PWEM or ISM algorithm to obtain an initial estimate of the electric field on the evaluation plane. Second, this initial estimate is refined through a deep learning model to yield a more accurate field distribution on the evaluation plane. From the reconstructed electric and magnetic fields, the spatially averaged power density (sPD) over an area $A$ is calculated as
	\begin{equation}
		\label{spd}
		sPD=\frac{1}{2A}\iint_A\mathfrak{R}\left[\boldsymbol{E}\times\boldsymbol{H}^*\right]\cdotp\boldsymbol{n}_AdA
	\end{equation}
	where \textbf{\textit{E}} and \textbf{\textit{H}} are the complex electric and magnetic fields on the evaluation plane, $(\cdot)^*$ denotes complex conjugation, $\mathfrak{R}(\cdot)$ extracts the real part, and $\textbf{\textit{n}}_A$ is the unit normal vector to the integration area. Two integration areas are considered, $1 \mathrm{cm}^{2}$ ($sPD_{1\mathrm{cm}^{2}}$) and $4 \mathrm{cm}^{2}$ ($sPD_{4\mathrm{cm}^{2}}$).
	
	\subsection{Plane-Wave Expansion Method}
	The PWEM reconstructs the field on the evaluation plane by decomposing the measured plane field into a superposition of plane waves and propagating each spectral component toward the desired plane. The process involves three main steps:
	
	(1) Spatial spectrum calculation: The 2D Fourier transform of the electric field measured on the plane at $z_{mea}$ yields the spatial spectrum
	\begin{equation}
		\label{A_mea}
		A(k_x,k_y,z_\mathrm{mea})=\iint \textbf{\textit{E}}(x,y,z_\mathrm{mea})e^{i(k_xx+k_yy)}dxdy
	\end{equation}
	
	(2) Spectral propagation: The spectrum is multiplied by the propagation operator to account for the phase shift between the measurement and evaluation planes, \emph{i.e.},
	\begin{equation}
		\label{A_rec}
		A(k_x,k_y,z_{\mathrm{eva}})=A(k_x,k_y,z_{\mathrm{mea}})e^{ik_z(z_{\mathrm{eva}}-z_{\mathrm{mea}})}
	\end{equation}
	
	(3) Field reconstruction: The inverse 2D Fourier transform of the propagated spectrum gives the reconstructed field
	\begin{equation}
		\label{E_rec}
		\textbf{\textit{E}}(x,y,z_{\mathrm{eva}})=\frac{1}{4\pi^2}\iint A_{\mathrm{eva}}e^{i(k_xx+k_yy)}\mathrm{d}k_x\mathrm{d}k_y
	\end{equation}
	where $k_x$, $k_y$ and $k_z$ are the x, y and z component of wave vector, respectively with the wavenumber k defined as $k_{x}^{2}+k_{y}^{2}+k_{z}^{2}=k^{2}$.

	This process can be expressed in matrix form as
	\begin{equation}
		\label{E_rec_M}
		\textbf{\textit{E}}_\mathrm{eva}=\textbf{\textit{F}}^{-1}\cdot \textbf{\textit{P}}\cdot \textbf{\textit{F}}\cdot \textbf{\textit{E}}_\mathrm{mea}
	\end{equation}
	where \textbf{\textit{F}} is the Fourier transform, \textbf{\textit{P}} is the diagonal propagation operator, and $\textbf{\textit{E}}_\mathrm{eva}$ and $\textbf{\textit{E}}_\mathrm{mea}$ are the vectors of reconstructed and measured field, respectively. 
	
	The PWEM suffers from an ill-conditioned problem when reconstructing evanescent plane-wave components, \emph{i.e.}, $k_x^2+k_y^2>k^2$ and $z_{\mathrm{eva}}<z_{\mathrm{mea}}$. In such cases, $e^{ik_z(z_{\mathrm{eva}}-z_{\mathrm{mea}})}$ grows exponentially, amplifying noise and modeling errors, and thereby degrading reconstruction stability.
	
	\subsection{Inverse Source Method}
	According to the equivalence theorem, the electric field radiated by DUT can be represented in terms of equivalent electric and magnetic surface currents \textbf{\textit{J}} and \textbf{\textit{M}} on a virtual closed surface $S$,
	\begin{equation}
		\label{Er}
		\boldsymbol{E}\left(\boldsymbol{r}\right)=\iint_{S}
		\begin{bmatrix}
			{\boldsymbol{G}}_{EJ}\left(\boldsymbol{r},\boldsymbol{r}^{\prime}\right)\cdotp\boldsymbol{J}\left(\boldsymbol{r}^{\prime}\right) \\
			+{\boldsymbol{G}}_{EM}\left(\boldsymbol{r},\boldsymbol{r}^{\prime}\right)\cdotp\boldsymbol{M}\left(\boldsymbol{r}^{\prime}\right)
		\end{bmatrix}d\boldsymbol{r}^{\prime2}
	\end{equation}
	where ${\boldsymbol{G}}_{EJ}$ and ${\boldsymbol{G}}_{EM}$ are the dyadic Green’s functions \cite{liu2025computational} for electric and magnetic currents with the following definitions
	\begin{equation}
		\label{GEJ}
		{\boldsymbol{G}}_{EJ}\left(\boldsymbol{r},\boldsymbol{r}^{\prime}\right)=-j\frac{\omega\mu}{4\pi}\left({\boldsymbol{I}}+\frac{\nabla\nabla}{k^2}\right)\frac{e^{-jk|\boldsymbol{r}-\boldsymbol{r}^{\prime}|}}{|\boldsymbol{r}-\boldsymbol{r}^{\prime}|}
	\end{equation}
	
	\begin{equation}
		\label{GEM}
		{\boldsymbol{G}}_{EM}\left(\boldsymbol{r},\boldsymbol{r}^{\prime}\right)=-\frac{1}{4\pi}\nabla\times\boldsymbol{I}\frac{e^{-jk|\boldsymbol{r}-\boldsymbol{r}^{\prime}|}}{|\boldsymbol{r}-\boldsymbol{r}^{\prime}|}
	\end{equation}
	where $\boldsymbol{I}$ is the identity matrix.
	
	Discretizing \eqref{Er} yields a linear system \textbf{\textit{b}} = \textbf{\textit{Ax}}, where  \textbf{\textit{b}} contains measured fields and \textbf{\textit{x}} represents the equivalent currents. Once  \textbf{\textit{x}} is determined, the field on any plane can be computed by forward evaluation of \eqref{Er}.
	
	ISM is also ill-posed, \emph{i.e.}, small perturbations in measurement data can cause large oscillations in the reconstructed solution. This sensitivity arises from the large condition number of \textbf{\textit{A}}, making solutions highly noise-sensitive.
	
	\subsection{Deep Learning Enhancement}
	
	\begin{figure*}[!t]
		\centering
		\includegraphics[width=5.5in]{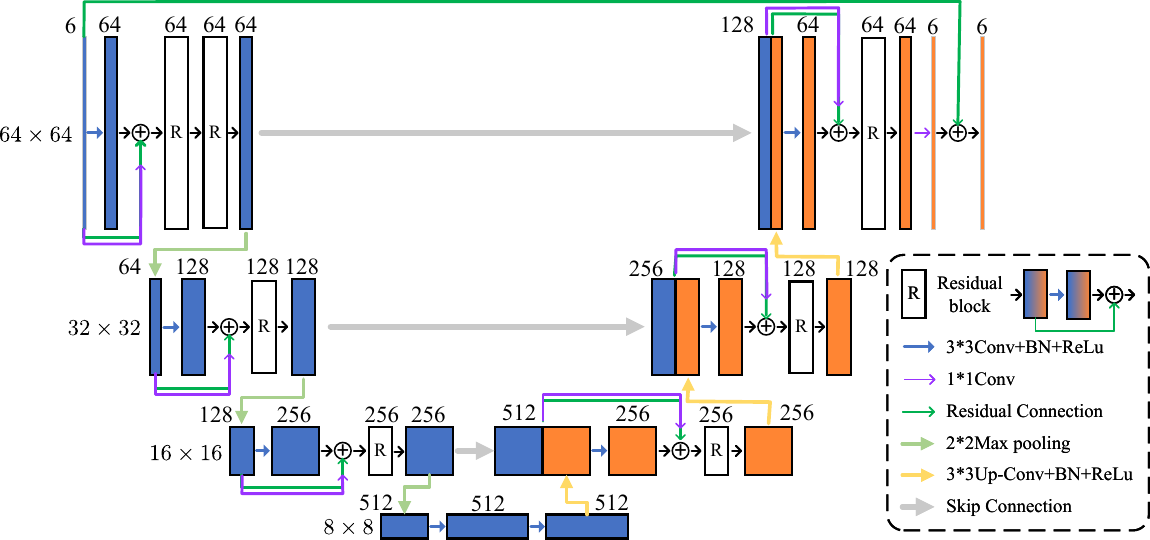}%
		\caption{The applied U-Net neural network architecture with residual connections, R-U-Net.}
		\label{R-U-Net}
	\end{figure*}
	
	U-Net is a convolutional neural network architecture, characterized by a symmetric encoder–decoder structure with skip connections between corresponding layers. This architecture is well-suited for problems requiring both global contextual understanding and precise spatial localization, making it a strong candidate for electromagnetic field reconstruction from sampled measurements \cite{ma2023inverse}.
	
	In this work, we employ a Residual U-Net (R-U-Net) architecture, which augments the standard U-Net with residual blocks. The residual block design enables the network to learn residual mappings rather than direct transformations, thereby alleviating the problems of gradient vanishing and gradient explosion that may occur in deep networks. The overall structure is illustrated in Fig.~\ref{R-U-Net}, where the encoder progressively reduces spatial resolution while capturing high-level features, and the decoder restores the resolution using transposed convolutions (up-convolutions) while integrating fine-grained details from the encoder through skip connections.
	
	We adopt a R-U-Net, the encoder of which contains four successive down-sampling stages and the decoder contains the corresponding four up-sampling stages. At each stage, a residual block consists of a sequence of convolution, batch normalization (BN), and ReLU activation layers, with a shortcut connection that adds the block input to its output. This residual design improves training efficiency and allows the network to be deeper without suffering performance degradation.
	
	The suitability of R-U-Net for the proposed field reconstruction framework can be explained by three key considerations:
	
	\begin{itemize}
		\item{IPD evaluation is most sensitive to regions with high electric field intensity, where accurate reconstruction is essential. The down-sampling operations in the encoder capture multi-scale features, enabling the network to model both broad spatial patterns and localized high-intensity regions effectively.}
		\item{The initial field estimates generated by PWEM or ISM and the desired accurate fields share significant structural features, especially in dominant field regions. Skip connections in U-Net allow low-level spatial information from the encoder to be directly passed to the decoder, ensuring that critical structural details are preserved and reducing reconstruction artifacts.}
		\item{While increasing network depth can improve the extraction of complex features, deeper architectures are prone to training instability and performance degradation. The residual learning mechanism in R-U-Net mitigates these risks by allowing the network to learn small corrective adjustments to the input rather than attempting to learn the entire mapping, thereby facilitating faster convergence and better generalization.}
	\end{itemize}
	
	Overall, R-U-Net effectively combines the advantages of encoder–decoder architectures with the stability benefits of residual learning. This makes it highly compatible with the proposed physics-informed reconstruction strategy, where the deep network refines the physically derived initial solution to achieve high-accuracy IPD estimation under various measurement conditions.
	
	\subsubsection{Dataset generation}

	\begin{figure*}[!t]
		\centering
		\includegraphics[width=5.5in]{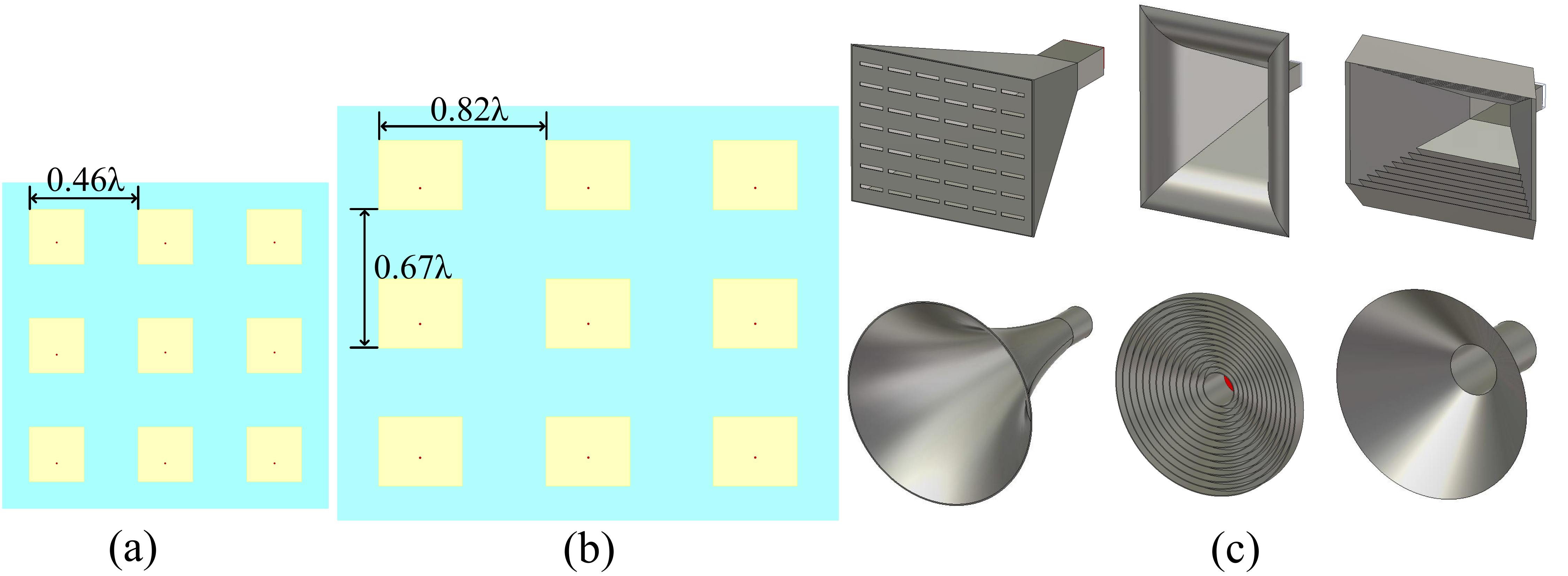}%
		\caption{Types of antennas used for data generation, (a) Square array, (b) rectangular array, and (c) horn antennas.}
		\label{antennas}
	\end{figure*}
	
	To ensure that the proposed reconstruction network can generalize across a wide range of antenna geometries and radiation characteristics, a diverse dataset is generated using full-wave electromagnetic simulations. The dataset includes two main antenna categories:
	\begin{itemize}
		\item{Patch array antennas: Two nine-port patch array configurations are considered—one with rectangular patch elements having unequal spacing in the $x-$ and $y-$directions (Fig.~\ref{antennas}(a)), and another with square patch elements having equal spacing in both directions (Fig.~\ref{antennas}(b)). In each case, the feed ports are randomly selected to excite the antenna, thereby producing a variety of radiation patterns and near-field distributions.}
		\item{Horn antennas: Several types of horn antennas are modeled, as shown in Fig.~\ref{antennas}(c). Variations include pyramidal and conical horns, as well as horns loaded with slot arrays and corrugated horns. Geometric parameters are systematically randomized to increase diversity, such as slot number, slot size, slot arrangement, corrugation order, and aperture dimensions. Notably, horns with slot arrays (referred to as “slot antennas” for brevity) produce near-field distributions that differ significantly from conventional horn antennas; therefore, they are treated as a separate class in subsequent analyses.}
	\end{itemize}
	
	The resulting dataset consists of 2,280 antenna configurations for training and 275 configurations for testing. This diversity ensures that the network is exposed to a wide variety of field distributions during training, enhancing its robustness to unseen antenna designs.
	
	\subsubsection{Training strategy}
	
	For each antenna configuration, the complex electric field is obtained at two planes parallel to the antenna aperture, the evaluation plane located 2 mm from the antenna representing the ultra-near region for IPD compliance evaluation and the measurement plane 22 mm from the antenna where field measurements can be performed with reduced probe–DUT interaction effects. The electric field on the measurement plane is processed using either PWEM or ISM to produce an initial estimate of the evaluation-plane field, which serves as the network input. The corresponding full-wave simulation result on the evaluation plane is used as the ground truth (label) for supervised training.
	
	We consider two training paradigms. The first one is named as independent training where separate networks are trained for each antenna type (\emph{i.e.}, array, slot, or horn) using only samples from that category. The other is called hybrid training where a single network is trained on the combined dataset containing all antenna types. The independent approach is expected to yield lower reconstruction error when tested on antennas similar to the training set, while the hybrid approach may offer better generalization to unseen antenna types and configurations.
	
	\begin{table}[htbp]
		\caption{Training parameters}
		\begin{center}
			\renewcommand{\arraystretch}{1.5}
			\begin{tabular}{ccccc}
				\hline
				Traindata& Array & Slot & Horn & Hybrid \\
				\hline
				Max epochs & 26 & 60 & 200 & 66\\ 
				Mini batch size & 8 & 16 & 64 & 64\\ 
				Initial learn rate & 0.001 & 0.001 & 0.001 & 0.001\\
				The learning rate decay factor & 0.5 & 0.5 & 0.5 & 0.5\\ 
				The learning rate decay period & 17 & 40 & 190 & 55\\ 
				\hline
			\end{tabular}
			\label{tab:1}
		\end{center}
	\end{table}
	
	The main training hyperparameters are summarized in Table~\ref{tab:1}. The networks are trained using the Adam optimizer, with an initial learning rate of $1\times 10^{-3}$ and a stepwise decay schedule. The mini-batch size and number of training epochs are selected according to antenna type and dataset size, as listed in the table. Both training strategies are evaluated in Section \ref{sec:Results} to determine the optimal approach for generalization and reconstruction accuracy.
	
	\section{Results and discussion}
	\label{sec:Results}
	
	The reconstruction performance of the proposed method is influenced by several factors, including the training strategy, the classical algorithm used to obtain the initial field estimate, the reconstruction distance, and the number of sampling points. Unless otherwise stated, results are based on antennas operating at 30 GHz, with both the measurement and evaluation planes sized at 10$\lambda$ × 10$\lambda$ and separated by 0.159$\lambda$. The evaluation plane is positioned 2 mm from the antenna, and the measurement plane is initially set at 22 mm.
	
	The following cases are defined for comparison:
	
	1) $E_{\mathrm{mea}}$: Electric field on the measurement plane.
	
	2) R-U-Net$_{\mathrm{I}}$: Field reconstruction based on a neural network trained with independent training strategy.
	
	3) R-U-Net$_{\mathrm{H}}$: Field reconstruction based on a neural network trained with hybrid training strategy.
	
	4) $E_{\text{PWEM}}$: Electric field on the evaluation plane obtained directly from the PWEM algorithm.
	
	5) $E_{\text{ISM}}$: Electric field on the evaluation plane obtained directly from the ISM algorithm.
	
	6) R-U-Net$_{\mathrm{PWEM}}$: Field reconstruction based on a neural network which takes PWEM solution as input.
	
	7) R-U-Net$_{\mathrm{ISM}}$: Field reconstruction based on a neural network which takes ISM solution as input.
	
	8) Truth: Ground truth.
	
	\subsection{Electric field reconstruction accuracy}
	
	Reconstruction error is quantified using the relative error (RE):
	\begin{equation}
		\label{RE}
		\mathrm{RE(\%)}=\frac{\left\|\left|{X}_\text{pre}\right|-\left|{X}_\text{ref}\right|\right\|_2}{\left\|\left|{X}_\text{ref}\right|\right\|_2}\times100
	\end{equation}
	where ${X}$ may represent the electric field, IPD, or peak spatially averaged IPD (psIPD). The subscript ``pre" denotes the predicted value and ``ref" being the ground truth.
	
	\subsubsection{Effect of training strategy}
	
	\begin{figure}[!t]
		\centering
		\includegraphics[width=0.7\linewidth]{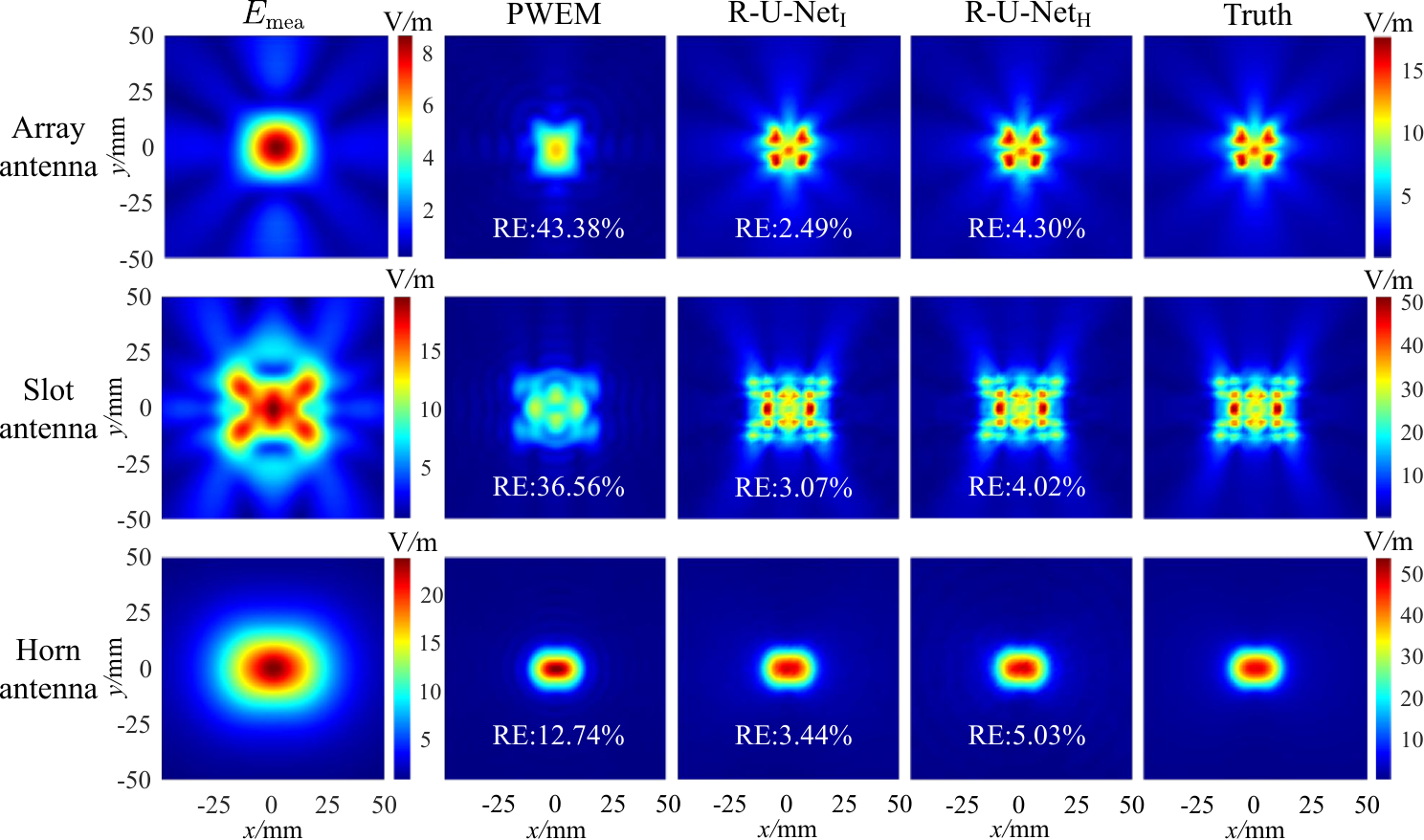}
		\caption{Comparison of predicted electric fields obtained by different training methods for three types of antennas. }
		\label{contrast_IH}
	\end{figure}
	
	\begin{table}[htbp]
		\caption{Average predication relative error with two training strategies}
		\begin{center}
			\renewcommand{\arraystretch}{1.5}
			\begin{tabular}{ccccccc}
				\hline
				&\multicolumn{3}{c}{Independent training} & \multicolumn{3}{c}{Hybrid training}\\
				\hline
				Antenna type& Array & Slot & Horn & Array & Slot & Horn\\\hline
				RE(mean) & 2.29\% & 3.80\% & 3.18\% & 3.79\% & 4.50\% & 4.44\%\\ 
				\hline
			\end{tabular}
			\label{tab:2}
		\end{center}
	\end{table}
	
	Figure~\ref{contrast_IH} compares reconstructed electric fields for three representative antennas (array, slot, and horn) using independent and hybrid training strategies. The Antennas are randomly selected from their respective test datasets and the initial estimate is obtained from PWEM.  For the same antenna type, independent training yields lower RE than hybrid training. This trend is consistent in the full test dataset, as shown in Table~\ref{tab:2}.

	\begin{figure}[!t]
		\centering
		\includegraphics[width=0.7\linewidth]{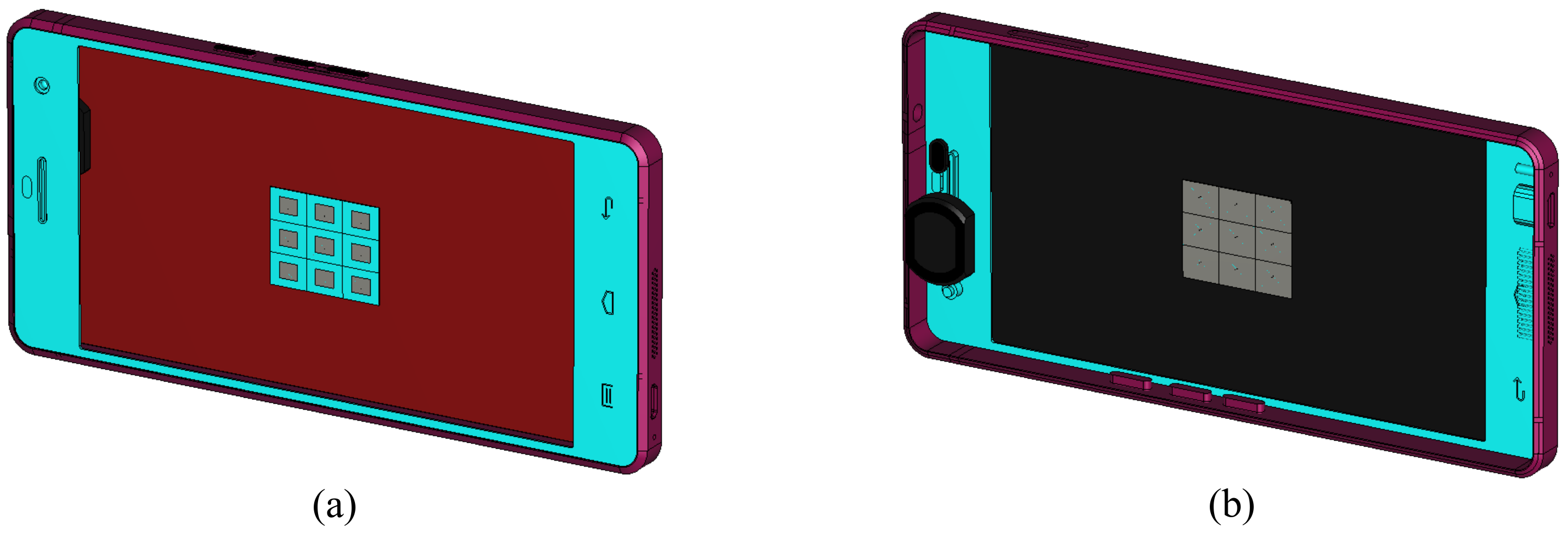}
		\caption{ (a) Front view (screen hidden) and (b) back view (backplate hidden) of the studied phone model.}
		\label{phone}
	\end{figure}
	
	\begin{figure}[!t]
		\centering
		\includegraphics[width=0.7\linewidth]{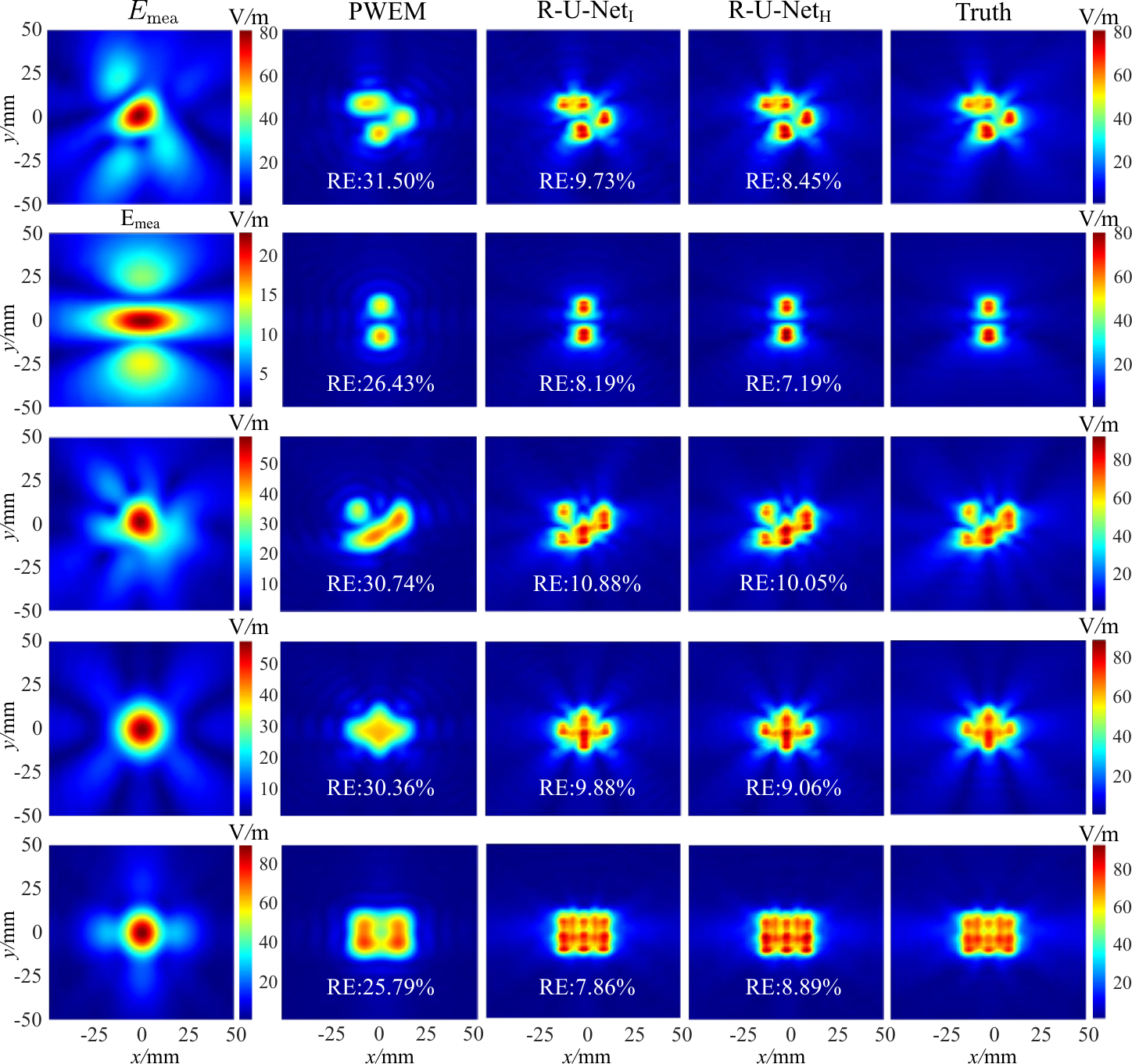}
		\caption{Comparison of predicted electric fields obtained by different training methods for phone with array antennas.}
		\label{phoneE}
	\end{figure}
	
	Although independent training improves accuracy for known antenna types, its generalization ability is limited. To assess generalization, an array antenna is embedded in a mobile phone model, as sketched in Fig.~\ref {phone}, to simulate realistic device conditions. Five antennas are randomly selected for simulation, and the measurement-plane fields are input to the two networks. Figure~\ref{phoneE} shows that hybrid training generally outperforms independent training in this scenario, producing lower RE despite the structural differences between training and test geometries. Consequently, the hybrid training strategy is adopted for subsequent experiments to ensure robustness across diverse use cases.
	
	\subsubsection{Effect of classical algorithm selection}
	
	\begin{figure*}[!t]
		\centering
		\includegraphics[width=0.9\linewidth]{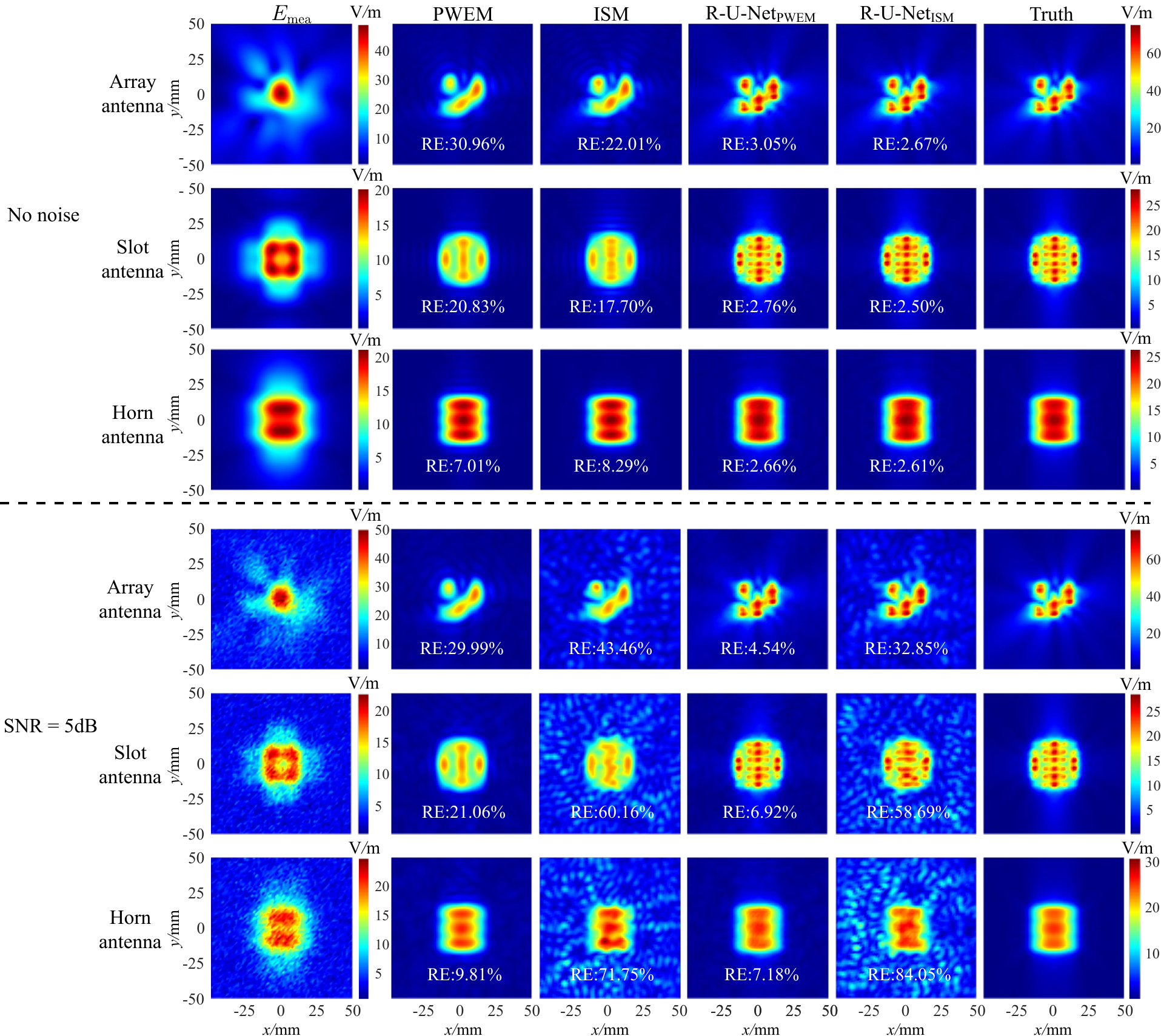}%
		\caption{Reconstructed electric field on the evaluation plane from PWEM, ISM and R-U-Net considering variation of noise level and antenna type.}
		\label{fig:contrast_noise}
	\end{figure*}
	
\begin{figure}[!t]
		\centering
		\includegraphics[width=2.6in]{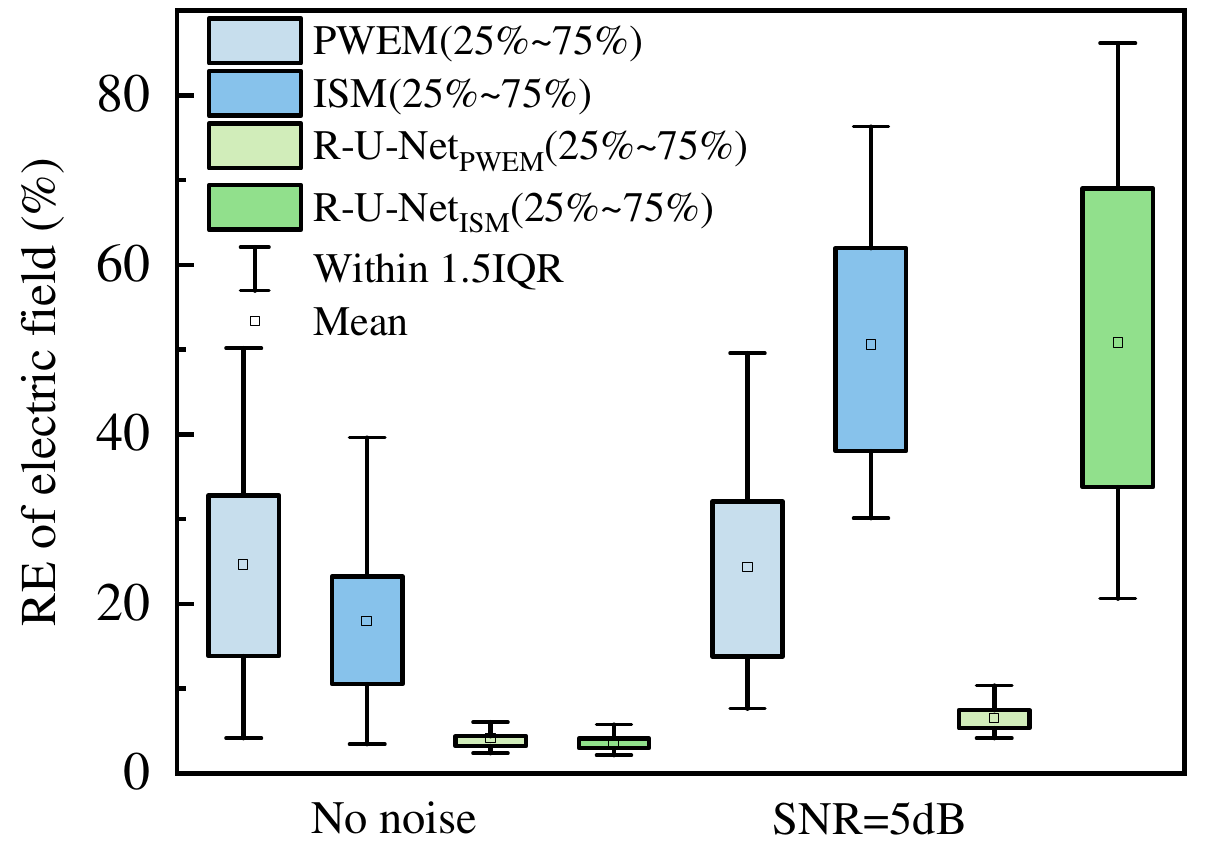}%
		\caption{Distribution of reconstruction errors for the applied four methods when reconstructing samples in the test dataset with different noise levels.}
		\label{fig:bpfourmethod}
	\end{figure}

	The initial evaluation-plane field can be derived from either PWEM or ISM before being refined by the network. Figure~\ref{fig:contrast_noise} compare results for four methods—PWEM, ISM, R-U-Net$_{\mathrm{PWEM}}$, and R-U-Net$_{\mathrm{ISM}}$) under noise-free and noisy conditions (SNR = 5 dB), respectively. In the absence of noise, all four methods capture the general field distribution, but PWEM and ISM tend to underestimate the amplitude of electric field, particularly for array and slot antennas. Both deep-learning-augmented methods achieve high accuracy, with R-U-Net$_{\mathrm{PWEM}}$ and R-U-Net$_{\mathrm{ISM}}$ producing nearly indistinguishable results from the truth. Under noisy conditions, ISM’s performance degrades significantly, which also affects R-U-Net$_{\mathrm{ISM}}$. In contrast, PWEM and R-U-Net$_{\mathrm{PWEM}}$ remain relatively stable, with the latter achieving the best balance between accuracy and robustness. The statistical results given in Fig.~\ref{fig:bpfourmethod} confirmed the above conclusion. Given these results, PWEM is selected as the preferred initial-value algorithm for subsequent analysis.
	
	\subsubsection{Effect of reconstruction distance}
	
	\begin{figure}[!t]
		\centering
		\includegraphics[width=2.6in]{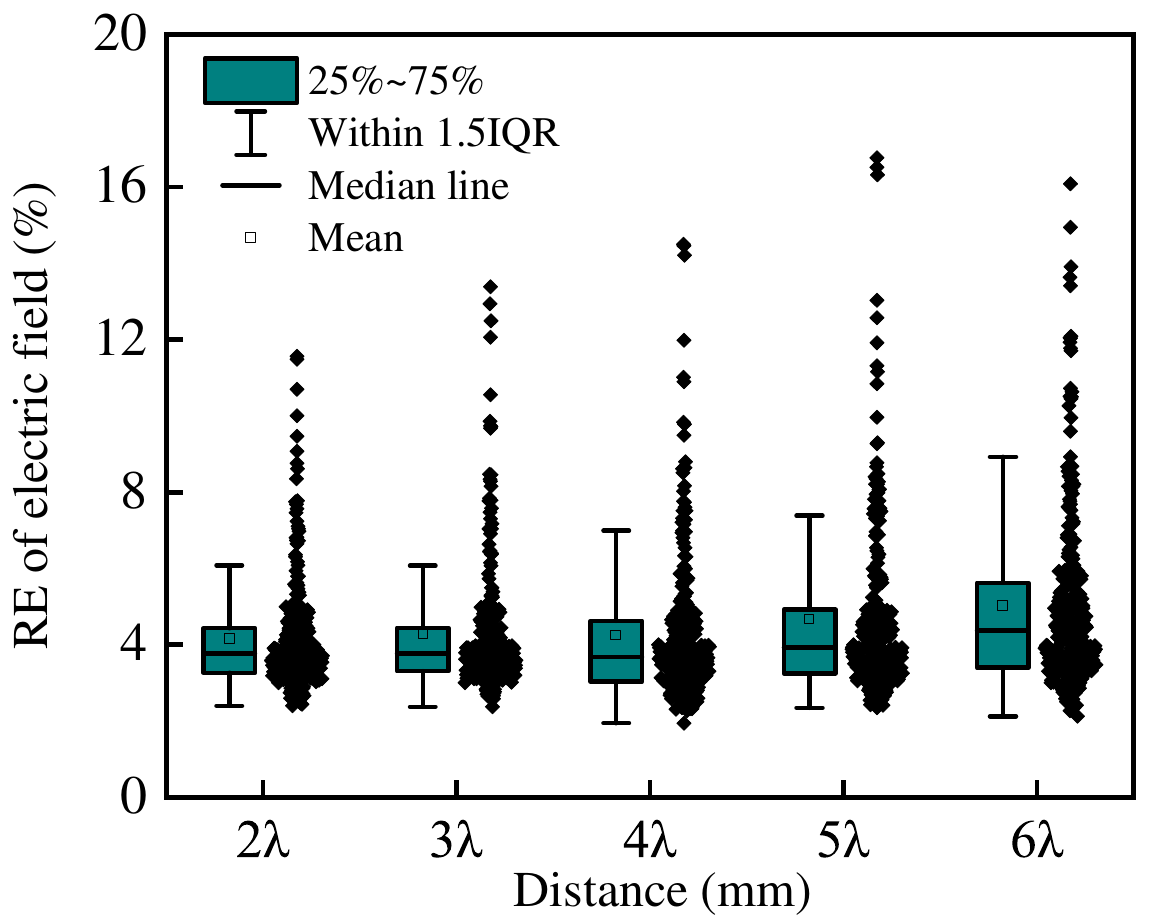}%
		\caption{Reconstruction errors when reconstructing samples in the test dataset with different evaluation distances using solvers from R-U-Net$_{\mathrm{PWEM}}$.}
		\label{fig:distance}
	\end{figure}
	
	To examine the effect of reconstruction distance, networks are trained for distances ranging from 
	2$\lambda$ to 6$\lambda$. Figure~\ref{fig:distance} shows the RE distribution across the test dataset for each case. As expected, larger distances lead to greater field attenuation and loss of high-frequency spatial components, increasing reconstruction difficulty. Nevertheless, even at 6$\lambda$, 75\% of the test antennas exhibit RE below 6\%, confirming the method’s stability over a broad range of distances. 
	
	\subsubsection{Effect of sampling density}
	
	\begin{figure}[!t]
		\centering
		\includegraphics[width=2.6in]{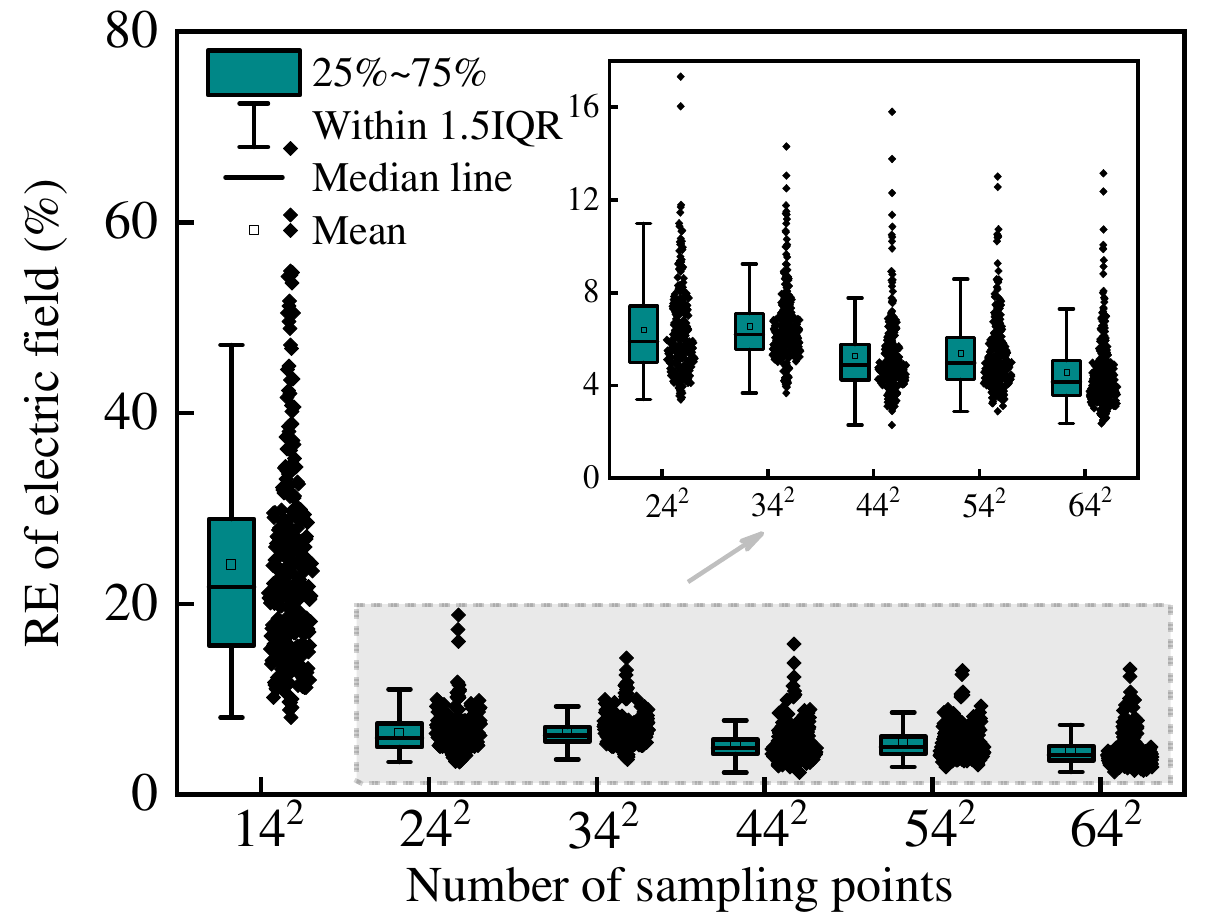}%
		\caption{Reconstruction errors for various numbers of sampling points.}
		\label{SP}
	\end{figure}
	
	The number of sampling points on the measurement plane directly impacts the quality of the initial PWEM reconstruction. Since the network’s input and output are $64\times 64$ matrices, fields sampled at lower resolution are first interpolated to this size before PWEM processing. Figure~\ref{SP} shows that at $14\times 14$ sampling, RE increases sharply, reaching 21.8\% on average. Increasing the sampling to $24\times 24$ dramatically improves performance, with 75\% of antennas achieving RE below 8\%. Further increases in sampling density continue to reduce RE, demonstrating that the method benefits from finer measurement resolution. In the following analysis, $64\times 64$ samples are collected from the measurement plane and used for the field reconstruction.
	
	\subsection{IPD distribution and psIPD evaluation}
	
	\begin{figure}[!t]
		\centering
		\includegraphics[width=3.3in]{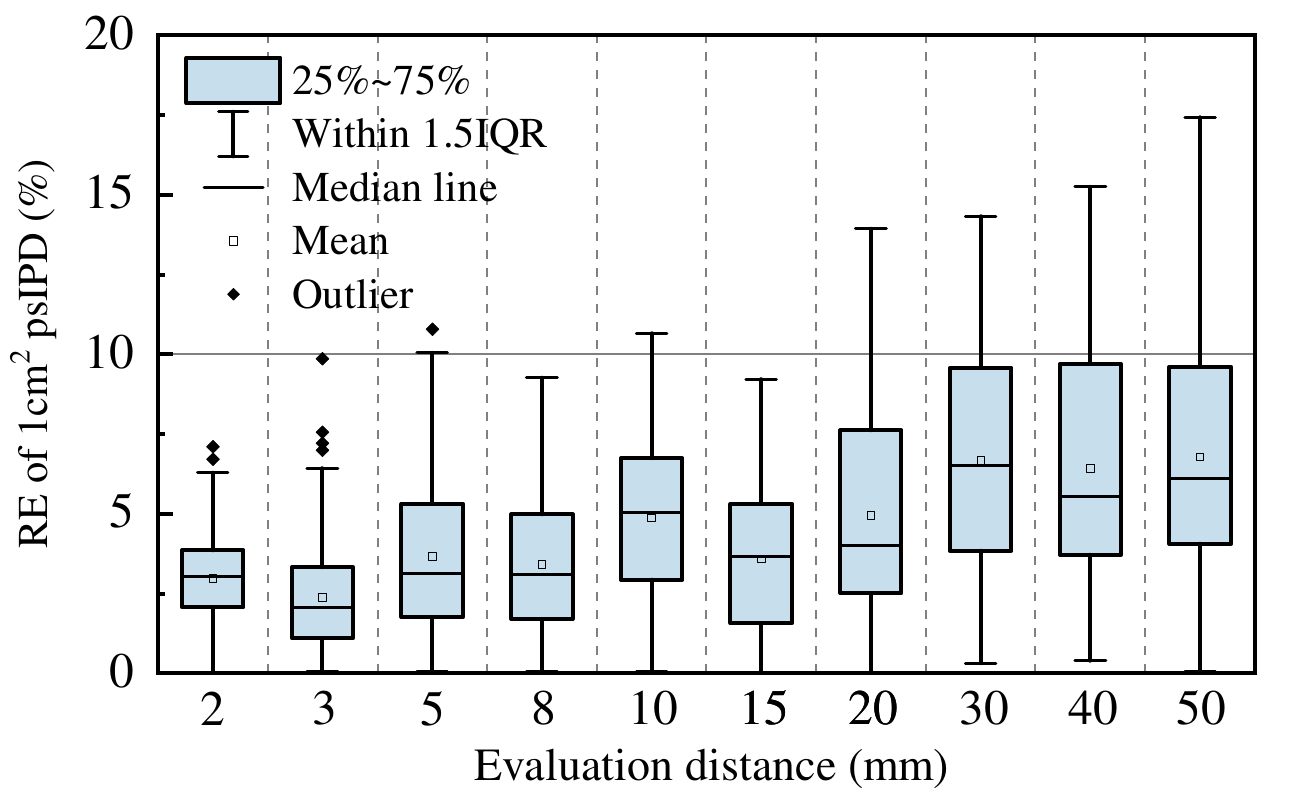}%
		\caption{Relative error of predicted $\mathrm{1cm^2}$ psIPD for samples in the test dataset with different evaluation distances.}
		\label{REpd}
	\end{figure}
	
	\begin{figure*}[!t]
		\centering
		\includegraphics[width=\linewidth]{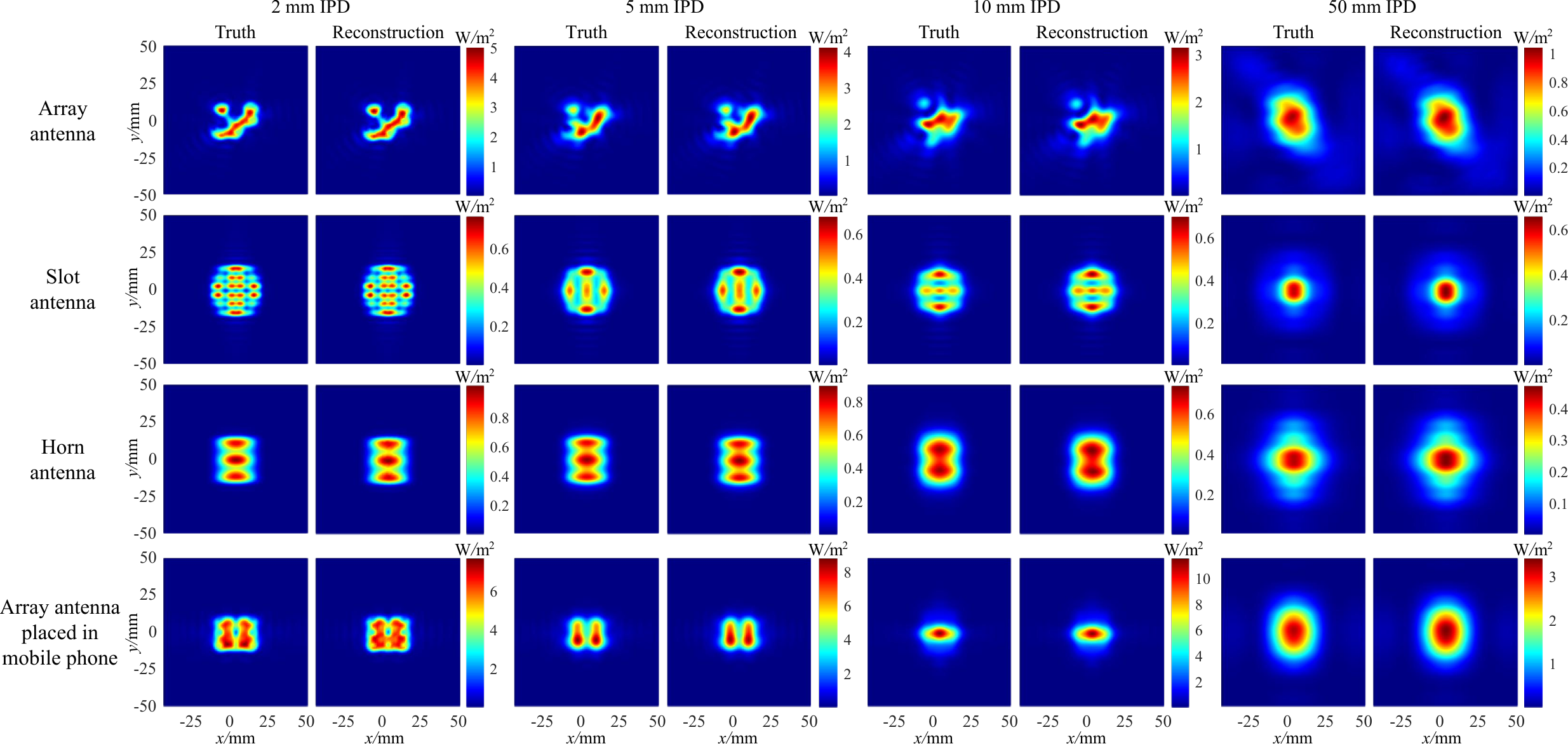}%
		\caption{Representative examples of estimated IPD distribution based on reconstructed electric field on the evaluation plane $z=2$mm, while electric fields on the other planes retrieved by PWEM. Four types of antennas are considered here for comparisons.}
		\label{IPD}
	\end{figure*}
	
	\begin{table}[htbp]
		\caption{Relative error (RE) and absolute error (AE) of predicted $\mathrm{1cm^2}$ psIPD for the four antenna types}
		\begin{center}
			\renewcommand{\arraystretch}{1.5}
			\begin{tabular}{cccccc}
				\hline
				\multirow{2}{*}{\begin{tabular}[c]{@{}c@{}}Antenna type\end{tabular}}&\multirow{2}{*}{\begin{tabular}[c]{@{}c@{}}Error\end{tabular}} & \multicolumn{4}{c}{Evaluation distance(mm)} \\  &  & 2 & 5 & 10& 50 \\\cline{1-6}
				\multirow{2}{*}{Array antenna}& RE(\%)& 3.35 & 2.49 & 1.98 & 3.74 \\ 
				& AE & 0.0975 & 0.0615 & 0.0456 & 0.0332 \\\cline{1-6}
				\multirow{2}{*}{Slot antenna} & RE(\%)& 0.83   & 7.45& 7.19 & 6.29\\ 
				& AE & 0.0040 & 0.0323 & 0.030  & 0.0343 \\\cline{1-6}
				\multirow{2}{*}{Horn antenna} & RE(\%)& 2.35  & 0.69  & 3.18  & 5.54 \\ 
				& AE & 0.0189 & 0.0052 & 0.0207 & 0.0235 \\\cline{1-6}
				\multirow{2}{*}{\begin{tabular}[c]{@{}c@{}}Array antenna placed\\in mobile phone\end{tabular}}& RE(\%)& 8.49   & 4.76   & 4.68   & 5.29   \\ 
				& AE& 0.1759 & 0.1049 & 0.2230 & 0.1719\\
				\hline
			\end{tabular}
			\label{tab:3}
		\end{center}
	\end{table}
	
	The proposed method’s ability to reconstruct IPD distributions is validated by comparing the estimation error of $1 \text{cm}^2$ psIPD across the test dataset at different distances. Although the network is trained to predict fields at 2 mm, PWEM enables transformation of the reconstructed field to other planes (e.g., 5 mm, 10 mm, and 50 mm).  As shown in Fig.~\ref{REpd}, when the evaluation distance is smaller than 20 mm, the estimated psIPD has RE$<$10\%. When the distance is larger, due to smaller values of psIPD, the obtained RE is higher even the absolute error is still small. 
	
	Figure~\ref{IPD} presents the IPD distributions for four representative antennas—array, slot, horn, and a phone-integrated array at multiple distances. In all cases, the reconstructed distributions accurately capture the focal regions and energy concentration patterns, confirming the method’s capability to assess both peak spatial-average IPD and IPD spatial distribution of exposure in the near-field region. The associated predicted relative error (RE) and absolute error (AE) are summarized in Table~\ref{tab:3}, where all RE values are below 10\%.
	
	\section{Uncertainty analysis of measurement}
	
	\begin{table*}[htbp]
		\caption{Description and distribution of uncertainty variables.}
		\begin{center}
			\renewcommand{\arraystretch}{1.5}
			\begin{tabular}{ccc}
				\hline
				Variable & Description & Distribution \\
				\hline
				$x_p,y_p,z_p$ & Cartesian coordinates of probe position& ${(x/y/z)}_{p}^\text{ref}+ \mathcal{U}(-u,u), u=0.2,0.4,0.6,0.8,1.0\mathrm{(mm)}$\\ \hline
				$|E|$ & Amplitude of electric field & $ |E|^\text{ref}+|E|^\text{ref}/\mathrm{SNR}, \mathrm{SNR}=5,10,15,20,25\mathrm{(dB)}$\\ \hline
				$\angle{E}\mathrm{(deg)} $& Phase of electric field &  $\angle{E}^\text{ref}+\mathcal{N}(0,\sigma), \sigma=5,10,15,20,25\mathrm{(deg)}$\\ 
				\hline	
			\end{tabular}
			\label{tab:4}
		\end{center}
	\end{table*}
	
	For the proposed reconstruction method to deliver reliable IPD evaluations, the quality of the measured electric field data must be sufficient to ensure accurate reconstruction. In practical measurement systems, several factors can degrade accuracy, including probe position errors, mutual coupling between probes, amplitude and phase uncertainties, and finite sampling resolution. This section quantifies the impact of these factors using Monte Carlo (MC) simulations, following the parameter definitions in Table~\ref{tab:4}. Here, $\mathcal{U}(a,b)$ denotes a uniform distribution over $[a,b]$, and $N(\mu,\sigma)$ denotes a normal distribution with mean $\mu$ and standard deviation $\sigma$.
	
	\subsection{Contribution of position offset uncertainty }
	\label{subsec:positionUncertainty}
	
	\begin{figure}[!t]
		\centering
		\includegraphics[width=\linewidth]{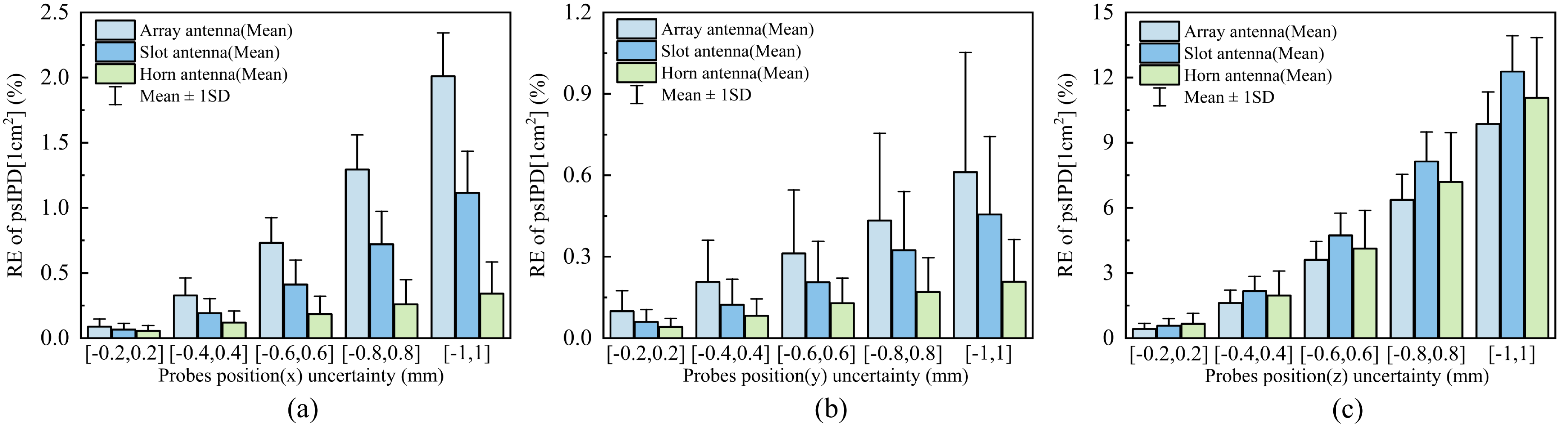}
		\caption{Relative error of $\mathrm{1cm^2}$ psIPD with different position offset along the (a) $x$ axis, (b) $y$ axis, or (c) $z$ axis.}
		\label{fig:positionUncertainty}
	\end{figure}
	
	Although probe arrays are calibrated before measurements, mechanical tolerances and alignment errors inevitably lead to deviations from nominal positions. Such errors can distort the measured field distribution.
	
	The effect of probe position error is evaluated by introducing uniformly distributed random offsets in the range [-$u$ mm, $u$ mm] to the true coordinates of each probe in the array. Examples when $u\in\{0.2, 0.4, 0.6, 0.8, 1\}$ are studied for comparisons. The resulting perturbed measurement-plane fields are used to generate the input of R-U-Net$_{\mathrm{PWEM}}$ model to predict evaluation-plane fields, from which the 1 cm\textsuperscript{2} psIPD is computed. This process is repeated for 1,000 Monte Carlo trials, and the average RE is calculated for three antenna types (array, slot, and horn antenna).
	
	As shown in Fig.~\ref{fig:positionUncertainty}, psIPD errors grow with increasing position deviation. The sensitivity to $z$-axis offsets is notably higher than for $x$- or $y$-axis offsets, as the field decays rapidly along the propagation direction, while in-plane variations are smoother due to spatial continuity. For $x$- and $y$-axis offsets, array antennas are most affected, followed by slot antennas, with horn antennas being least sensitive. An exception occurs for $z$-axis offsets, where slot antennas exhibit the largest errors. When the position deviation in all directions is within [-0.8mm, 0.8mm] (1/12$\lambda$), the psIPD error remains below 10\%, indicating that probe placement tolerances can be relaxed to this level without significantly compromising accuracy.
	
	\subsection{Contribution of coupling effect uncertainty}
	Probe arrays enable rapid acquisition of spatial field data but are susceptible to mutual coupling between adjacent elements. This coupling can alter the measured field, especially when probe geometries and loading effects vary slightly among elements.
	
	Following \cite{LIU}, the coupling effect is modeled by considering only the eight nearest neighbors to each probe. The measured field at position $\boldsymbol{r}$ is expressed as
	\begin{equation}
		\label{coupling_effect}
		\boldsymbol{E}(\boldsymbol{r})=\boldsymbol{E}_\text{ref}(\boldsymbol{r})+\sum_{p_x=-1}^1\sum_{p_y=-1}^1
		\begin{bmatrix}
			c^{p_x,p_y}_{x,x} & c^{p_x,p_y}_{x,y} \\
			c^{p_x,p_y}_{y,x} & c^{p_x,p_y}_{y,y}
		\end{bmatrix}\boldsymbol{E}_\text{ref}(\boldsymbol{r})
	\end{equation}
	where ${p_x}$=${p_y}$=0 corresponds to the probe under concern. $\boldsymbol{E}_\text{ref}$ is a vector composed of x- and y- component of electric field without coupling disturbance. $c^{p_x,p_y}_{x,y}$ are the coupling coefficients. For simplicity, the same coupling matrix is assumed for all probes, with reference values taken from \cite{LIU}.
	
	\begin{figure}[!t]
		\centering
		\includegraphics[width=2.6in]{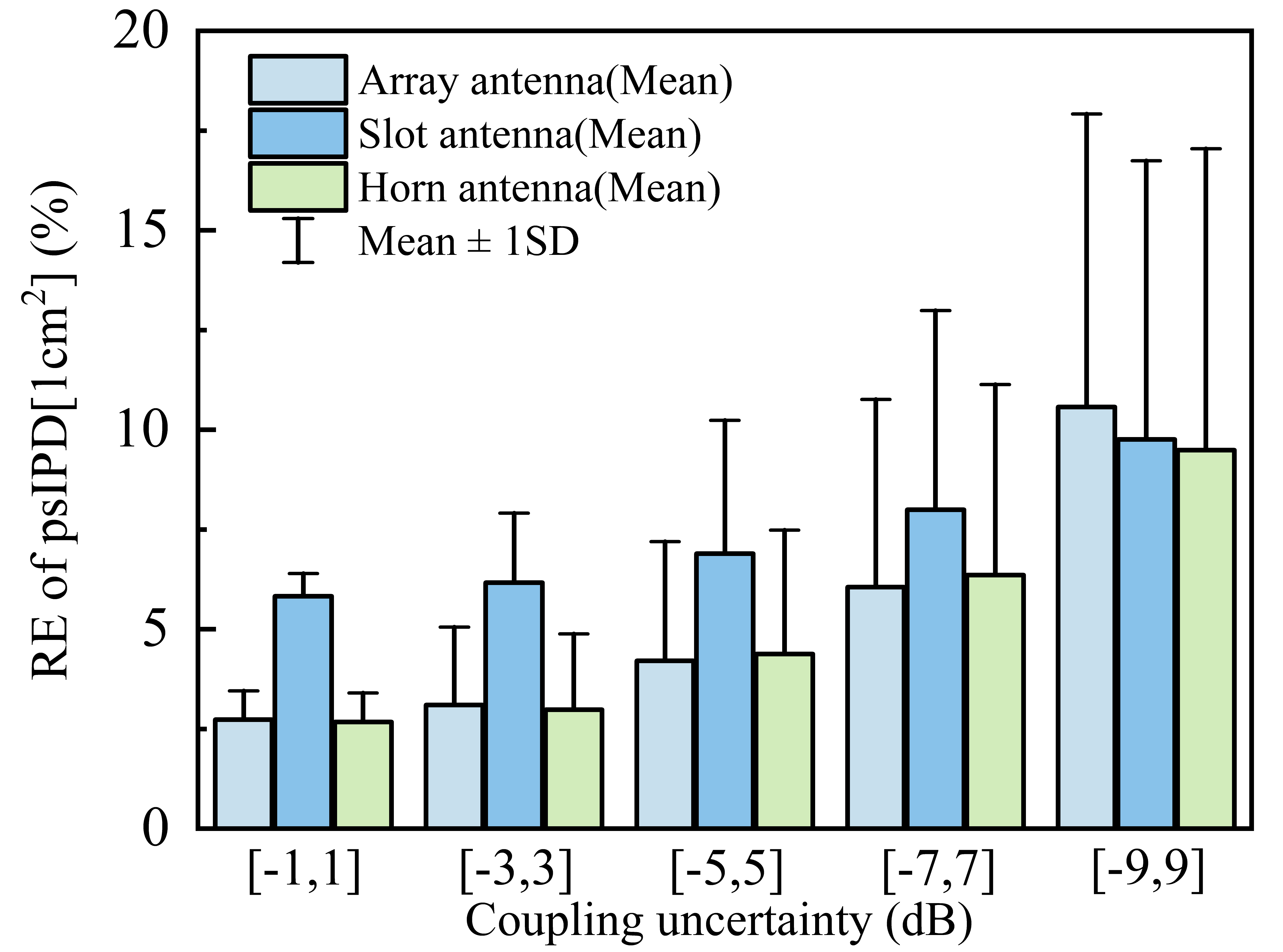}
		\caption{Relative error of $\mathrm{1cm^2}$ psIPD with different coupling coefficients.}
		\label{coupling}
	\end{figure}
	
	Among 1000 MC trials, coupling coefficients are perturbed by uniformly distributed random values in  [-$u$ dB, $u$ dB]. $u\in\{1, 3, 5, 7, 9\}$ is considered for analysis. Fig.\ref{coupling} shows that the sensitivity to coupling coefficient variation is greater than to position offset. However, if the coupling remains within $\pm$7 dB, the psIPD error stays below 10\%, indicating that moderate coupling levels are tolerable in practical systems.
	
	\subsection{Contribution of amplitude and phase uncertainty}
	
	\begin{figure}[!t]
		\centering
		\includegraphics[width=0.65\linewidth]{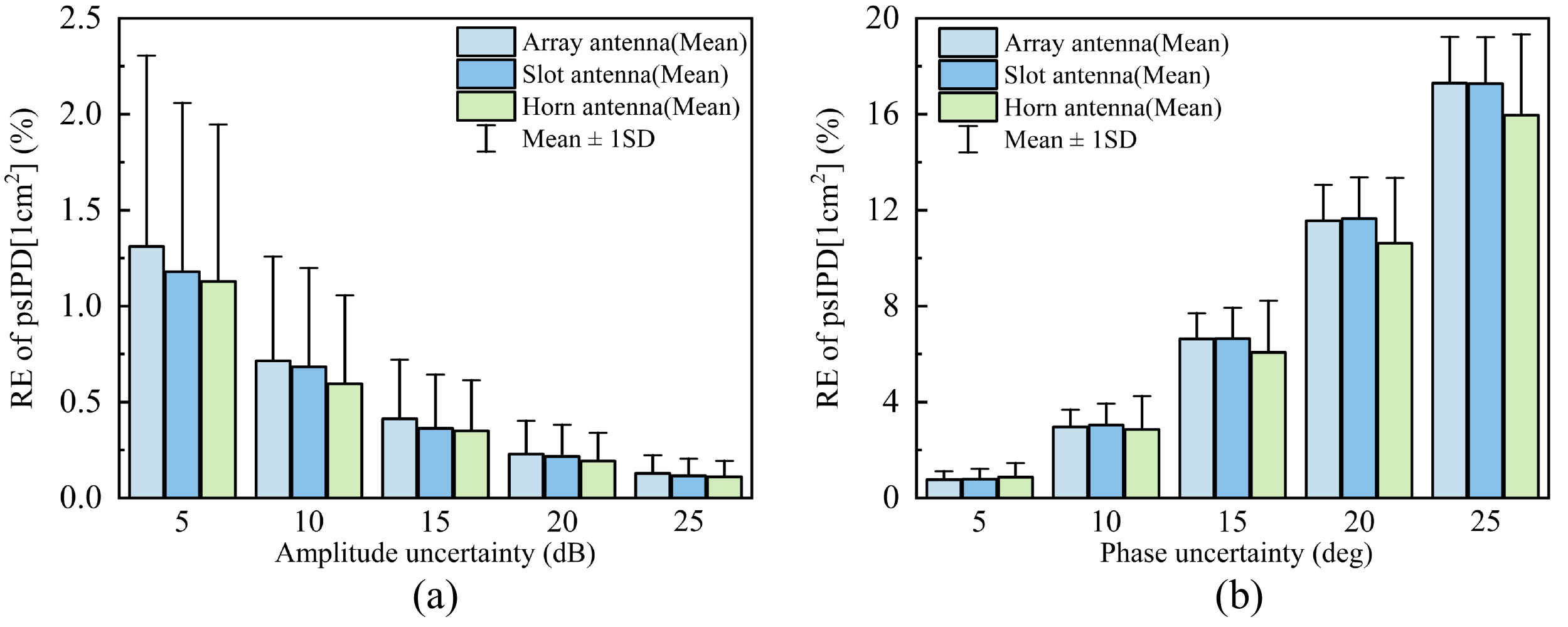}
		\caption{ Relative error of $\mathrm{1cm^2}$ psIPD considering uncertainties from (a) amplitude and (b) phase measurement. Amplitude uncertainty quantified by SNR and phase follows normal distribution with different standard deviations.}
		\label{fig:measurementUncertainty}
	\end{figure}
	
	Measurement noise can be modeled as uncertainties in both the amplitude and phase of the electric field. Amplitude noise is modeled by adding a normally distributed perturbation proportional to $|E_\text{ref}|$ with a standard deviation corresponding to the specified signal-to-noise ratio (SNR). Phase noise is modeled by adding zero-mean Gaussian noise with standard deviation $\sigma$ degrees. Fig.~\ref{fig:measurementUncertainty} shows psIPD error versus amplitude and phase noise levels, respectively, averaged over 1,000 MC trials. The method exhibits strong robustness to amplitude noise: even at SNR = 5 dB, psIPD errors remain below 2.5\% for all antenna types. In contrast, phase noise has a more pronounced effect. Errors remain under 8\% when $\sigma\le$15°, but increase rapidly beyond this threshold. Horn antennas are less affected by phase noise than arrays or slot antennas, suggesting greater inherent stability in their field distributions.
	
	The above analysis confirms that the proposed R-U-Net$_{\mathrm{PWEM}}$ method is robust to typical measurement uncertainties. Errors can be kept within acceptable limits if probe position tolerances, coupling levels, and noise performance are controlled to the ranges identified above.
	
	\section{Experimental validation}
	\label{sec:ExperimentalValidation}
	
	\begin{figure}[!t]
		\centering
		\includegraphics[width=2.6in]{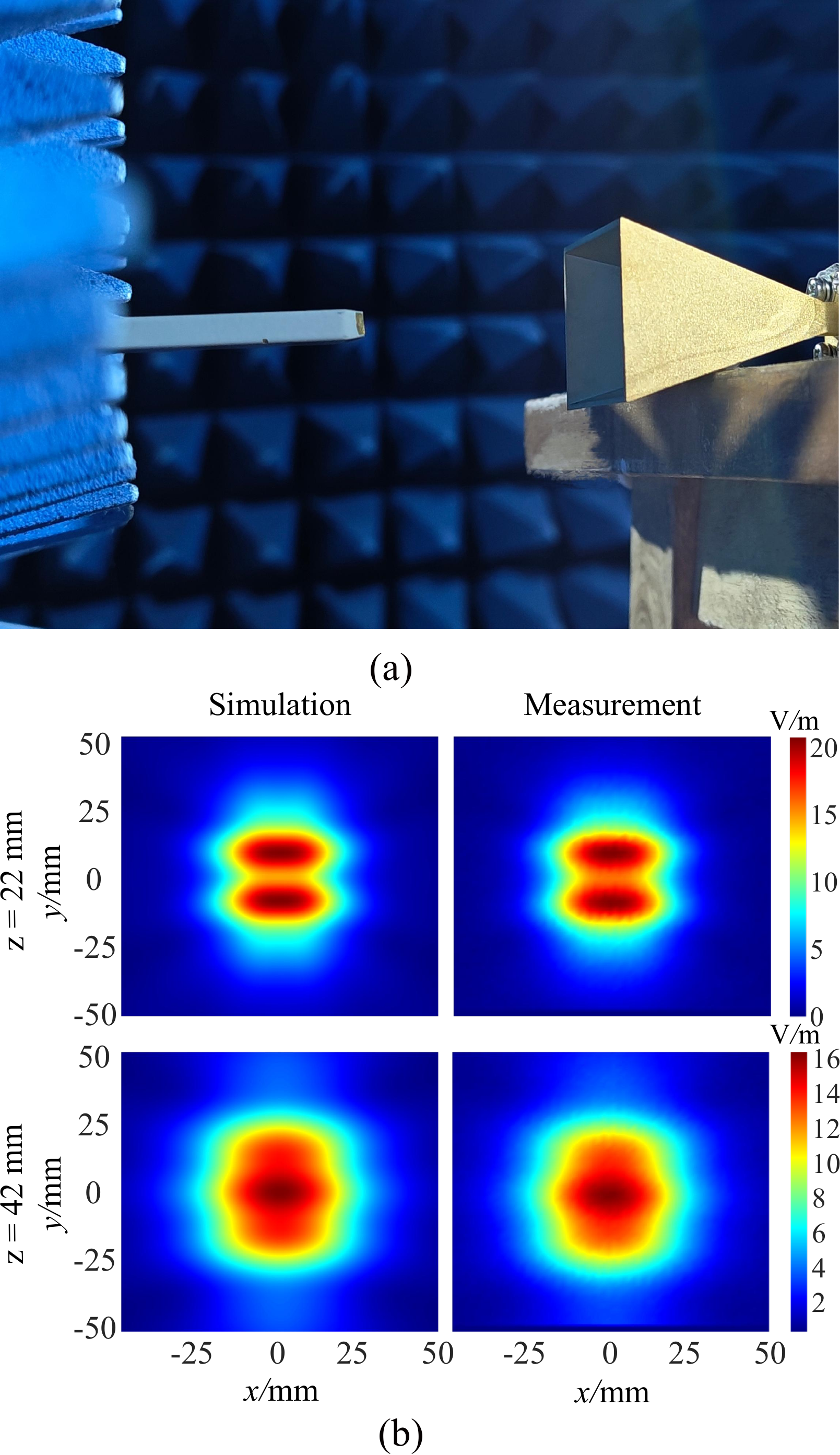}
		\caption{(a) Conical horn reference antenna defined in IEC/IEEE 63195. (b) Comparison between simulated and measured electric fields at $z$ = 22 mm and $z$ = 42 mm.}
		\label{Measurement}
	\end{figure}
	
	An experimental campaign was conducted to validate the effectiveness of the proposed reconstruction method using the conical horn reference antenna defined in the IEC/IEEE 63195 standard \cite{63195–1}, as shown in Fig.~\ref{Measurement}(a). The measurement setup consisted of a waveguide probe (TSB-320SGAH20K) and a vector network analyzer (Ceyear 3671G). The reference antenna operated at 30 GHz with a wavelength of $\lambda$ = 10 mm. The antenna aperture was positioned at $z$ = 0 mm, and electric fields were measured on two planes at $z$ = 22 mm and $z$ = 42 mm. Each measurement plane covered an area of $10\lambda\times 10\lambda$, sampled with a grid of $64\times 64$ points.
	
	The measured fields at $z$ = 22 mm and $z$ = 42 mm were used to reconstruct the fields at $z$ = 2 mm, corresponding to reconstruction distances of 20 mm (2$\lambda$) and 40 mm (4$\lambda$), respectively. These two distances were selected to evaluate whether the proposed framework conforms to the accuracy trends summarized in the  previous section. Due to the physical constraints of the measurement probe, direct acquisition of the electric field at $z$ = 2 mm was not feasible; therefore, simulated fields were used as the reference dataset for this plane.
	
	Figure~\ref{Measurement}(b) compares simulated and measured fields at the two measurement planes. Overall, strong consistency is observed, with only slight deviations in high-field regions. These discrepancies are primarily attributed to environmental noise in the measurement setup and mechanical tolerances in antenna fabrication.
	
	\begin{figure}[!t]
		\centering
		\includegraphics[width=4in]{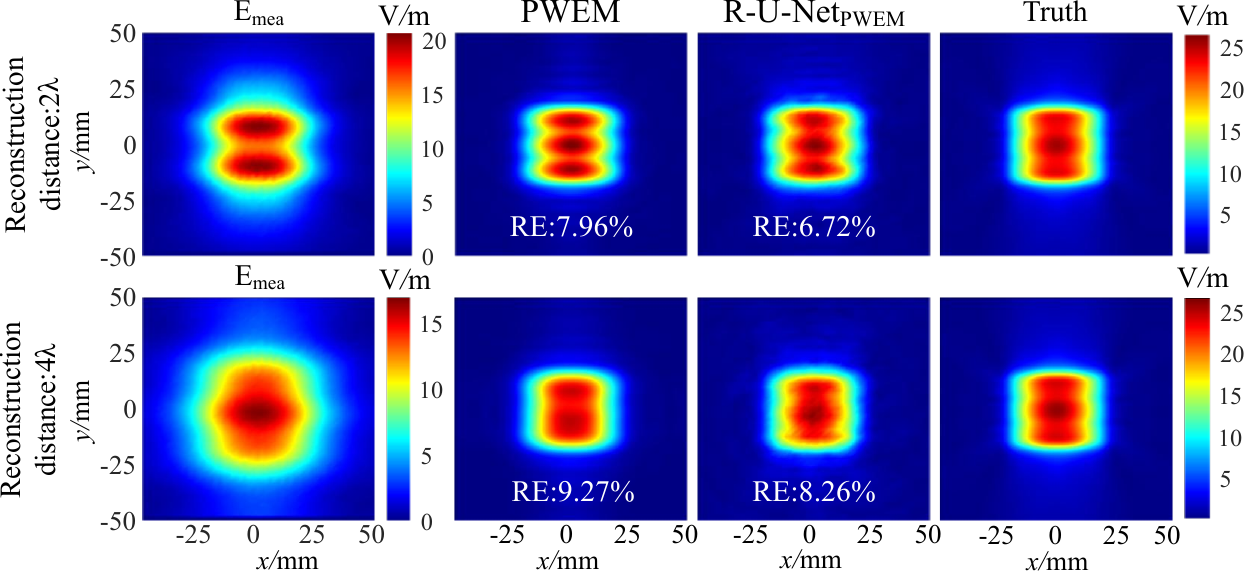}%
		\caption{Reconstructed electric fields using PWEM and R-U-Net$_{\mathrm{PWEM}}$ based on measured fields at $z$ = 22 mm (a) and $z$ = 42 mm (b), respectively. The reconstruction accuracy is quantified by relative error (RE).}
		\label{meaE}
	\end{figure}
	The measured data were subsequently applied to validate the R-U-Net$_{\mathrm{PWEM}}$ framework. As shown in Fig.~\ref{meaE}, the reconstructed fields obtained with R-U-Net exhibit consistently lower relative error (RE) than those reconstructed using PWEM alone, for both 
	$z$ = 22 mm and $z$ = 42 mm. While the improvement is not as dramatic as in simulation-only tests—due to the presence of measurement uncertainty—the accuracy gains are still substantial and confirm the method’s robustness under practical conditions.
	
	\begin{figure*}[!t]
		\centering
		\includegraphics[width=\linewidth]{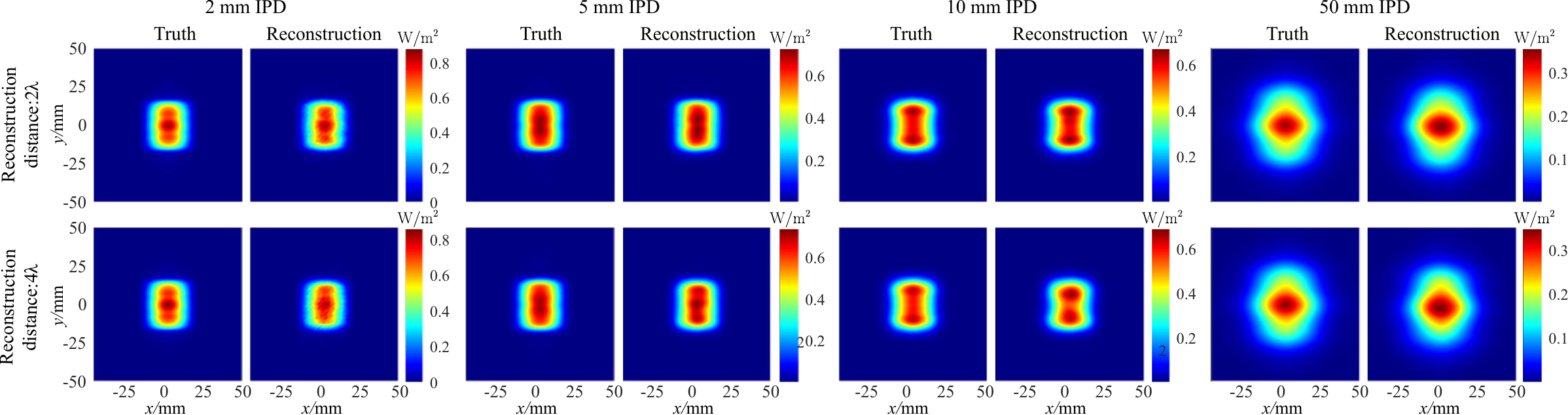}%
		\caption{Reconstructed incident power density (IPD) distributions at evaluation distances of 2 mm, 5 mm, 10 mm and 50 mm and the ground truth.}
		\label{meaIPD}
	\end{figure*}
	The reconstructed fields were further used to compute the incident power density (IPD). Figure~\ref{meaIPD} presents the IPD distributions at evaluation distances of 2 mm, 5 mm, 10 mm and 50 mm. In all cases, the reconstructed distributions closely match the simulated references, successfully capturing both the magnitude and spatial distribution of energy concentration.
	
	\begin{table}[!t]
		\caption{Relative and absolute errors of psIPD reconstructed from measured electric fields at different evaluation distances.}
		\begin{center}
			\renewcommand{\arraystretch}{1.5}
			\begin{tabular}{cccccc}
				\hline
				\multirow{2}{*}{\begin{tabular}[c]{@{}c@{}}Reconstruction\\distance\end{tabular}}&\multirow{2}{*}{\begin{tabular}[|c|]{@{}c@{}}Error\\ type\end{tabular}} & \multicolumn{4}{c}{Evaluation distance(mm)} \\ \cline{3-6} &  & 2 & 5 & 10& 50 \\
				\hline
				\multirow{2}{*}{2$\lambda$}& RE(\%)& 0.32 & 0.12 & 2.31 & 4.08 \\ \cline{2-6}
				& AE & -0.0024 & -0.0008 & 0.0137 & 0.0131 \\
				\hline
				\multirow{2}{*}{4$\lambda$} & RE(\%)& 1.60   & 1.12& 1.98 & 2.93\\ \cline{2-6}
				& AE & -0.121 & -0.0076 & 0.0117  & 0.0094 \\
				\hline
			\end{tabular}
			\label{tab:5}
		\end{center}
	\end{table}
	The relative and absolute errors of psIPD are summarized in Table~\ref{tab:5}. As observed, the RE values are all below 5\%, demonstrating the high reconstruction accuracy of the proposed IPD estimation scheme. Note that increasing evaluation distance may lead to larger RE (e.g., from 10 mm to 50 mm), although the absolute error (AE) keeps at small values.
	
	These experimental results confirm that the proposed R-U-Net$_{\mathrm{PWEM}}$
	method not only improves field reconstruction accuracy over conventional PWEM in realistic measurement scenarios but also provides reliable IPD estimates across multiple evaluation distances, thereby validating its effectiveness for practical exposure assessment.
	
	\section{Conclusion}
	This work presented a hybrid field reconstruction framework for accurate incident power density (IPD) evaluation of wireless devices working above 6 GHz, combining classical electromagnetic algorithms with deep learning. The proposed method first applies a physics-based algorithm, the plane wave expansion method (PWEM) or inverse source method (ISM), to generate an initial estimate of the electric field on the evaluation plane. This estimate is then refined using a Residual U-Net model trained on a diverse, full-wave-simulated dataset comprising array, slot, and horn antennas.
	
	Two training strategies were investigated. While independent training achieved lower reconstruction errors for antennas of the same type as the training set, hybrid training demonstrated superior generalization to unseen configurations, particularly in realistic device scenarios. Comparative analysis of classical algorithms showed that PWEM offers better robustness to measurement noise than ISM, making it the preferred choice for initial-value computation.
	
	Performance evaluations demonstrated that the proposed R-U-Net$_{\mathrm{PWEM}}$ approach significantly reduces reconstruction errors, achieving an average electric field RE of 4.57\% and psIPD RE of 2.97\% across the test dataset. The method maintained stability for reconstruction distances up to $6 \lambda$ and for sampling densities down to $24\times 24$ points. IPD distributions reconstructed from the 2 mm evaluation plane accurately captured focal regions at arbitrary distances via near–far transformation.
	
	A comprehensive uncertainty analysis quantified the effects of probe position deviation, inter-probe coupling, and amplitude/phase noise. Results indicated that the proposed method remains within 10\% psIPD error if probe offsets are within $\pm$0.8 mm, coupling coefficients within $\pm$7 dB, and phase noise below 15°. Based on the horn antenna in the IEC/IEEE 63195, experiments are carried out to verify the proposed method. The experiments show that the method is effective and can improve the reconstruction accuracy on the basis of the classical algorithm.
	
	Overall, the integration of classical field transformation with deep learning enables accurate, stable, and efficient IPD evaluation in the ultra-near field, addressing limitations of purely physics-based methods. The framework provides a practical pathway for reliable RF exposure compliance assessment of next-generation millimeter-wave wireless devices.
	
	\bibliographystyle{chicago}
	\bibliography{ref}
\end{document}